\documentclass[12pt,reqno]{amsart}
\allowdisplaybreaks[4]
\usepackage{graphicx} 
\usepackage{amssymb}
\usepackage{amsmath}
\usepackage{cite}

\numberwithin{equation}{section}
\makeatletter      
\@addtoreset{equation}{section}
\makeatother       

\begin{document}

\title{Additional symmetries of constrained  CKP and BKP hierarchies }

\author{Kelei Tian\dag, Jingsong He$^*$\ddag, Jipeng Cheng\dag \  and  Yi
Cheng \dag}
\dedicatory { \dag\ Department of Mathematics, University of Science and Technology of China,   China\\
\ddag\  Department of Mathematics, Ningbo University,  China
 }
\thanks{$^*$Corresponding author. email: hejingsong@nbu.edu.cn}
\begin{abstract}

  The additional symmetries of the constrained  CKP (cCKP)  and   BKP (cBKP)
  hierarchies are given by their actions on the Lax operators, and  their actions
on the eigenfunction and adjoint eigenfunction $\{\Phi_i,\Psi_i \}$
are presented  explicitly. Furthermore, we show that acting on the
space of the wave operator, $\partial_k^*$ forms new centerless
$W^{cC}_{1+\infty}$ and $W^{cB}_{1+\infty}$-subalgebra of centerless
$W_{1+\infty}$ respectively. In  order to define above symmetry
flows $\partial_k^*$ of the cCKP and cBKP hierarchies, two vital
operators $Y_k$ are introduced to revise the additional symmetry
flows of the CKP and BKP hierarchies.

\end{abstract}


 \maketitle

 PACS (2003). \ 02.30.Ik.





\section{Introduction}

The research on  Kadomtsev-Petviashvili (KP) hierarchy
\cite{dkjm,dl1}  is one of the most important  topics in the
development of the  theory of integrable systems. A specific
interesting aspect on this topic  is  additional
symmetry\cite{os1,asv1,asv2,asv3,dl3,dl4}. Additional symmetries are
 special symmetries which are not contained in the KP
hierarchy and do not commute  with each other. The additional
symmetry flows of the KP hierarchy form an infinite dimensional
algebra $W_{1+\infty}$\cite{asv2,dl4}. More recently, there are
several new results   about partition function in the matrix models
and Seiberg-Witten theory associated with additional symmetries,
string equation and Virasoro constraints of the KP hierarchy
\cite{Morozov1994,Morozov1998,Alexandrov, Mironov2008,Aratyn2003}.
In fact, there is a parallel research line on additional symmetries
of evolution equations and their Lie algebraic structures,
particularly for $1+1$ dimensional integrable equations \cite{ma1},
and the corresponding symmetries are called $\tau$ -symmetries and
related Lax operators are constructed pretty systematically
\cite{ma2,ma3}.

It is well known that we can get different sub-hierarchies of the KP
by different reduction conditions on Lax operator $L$. The first two
important sub-hierarchies are CKP hierarchy \cite{dkjm2} through a
restriction $L^*=-L$ and BKP hierarchy\cite{dkjm} through a
restriction $L^*=-\partial L
\partial^{-1}$. So the additional symmetries of the CKP (or BKP) hierarchy  have been constructed
\cite{he2007,Takasaki1993,tu1,tu2} and shown to be a new infinite
dimensional algebra $W^C_{1+\infty}$ (or $W^B_{1+\infty}$),  which
is subalgebra of $W_{1+\infty}$. The third sub-hierarchy is
constrained KP (cKP) hierarchy \cite{kss,chengyiliyishenpla1991,
chengyiliyishenjmp1995}, which is obtained by setting a special form
of Lax operator
 $L=\partial
+\sum _{i=1}^m \Phi_i \partial ^{-i}\Psi _i,$  where $ \Phi_i $ is
the eigenfunction and $\Psi _i $ is the adjoint eigenfunction of the
cKP hierarchy.  This is a relative new reduction condition and is
different from the reduction conditions of previous two
sub-hierarchies. Hence it is very natural to explore the additional
symmetries of the cKP hierarchy. However, additional symmetry flows
of the KP hierarchy break the special form of the Lax operator for
the cKP hierarchy, and so it is highly non-trivial  to find a
suitable form of additional symmetry flow for this sub-hierarchy. To
this end,  it is well defined\cite{aratyn1997} by means of a crucial
modification of the corresponding additional symmetry flows of the
KP hierarchy. Note that a new and complicated operator $X_k$ plays a
very important role in this procedure.

Furthermore, the fourth sub-hierarchy and fifth sub-hierarchy are
constrained CKP (cCKP)\cite{Loris1999} and constrained BKP (cBKP)
hierarchy\cite{Loris1999JMP}, which are obtained by putting the
combination of the previous mentioned reduction conditions, i.e.,
CKP  and cKP conditions, BKP and cKP conditions. Some results
associated  cCKP and cBKP hierarchies have already been reported,
for example, a Gramm-type $\tau$-function \cite{Loris1999} of the
cCKP hierarchy and a Pfaffian-form $\tau$-function
\cite{Loris1999JMP} of the cBKP are also constructed from the vacuum
solution, the dimensional reductions of the CKP and BKP hierarchies
are presented in \cite{Loris2001}. Moreover, determinant
representations of the $\tau$ function, which are generated by the
the gauge transformations, of the cCKP and cBKP hierarchies are
obtained \cite{HeWu}. However, for our best knowledge, there is no
any results on the additional symmetries of the cBKP and cCKP
hierarchies. So, we shall fill the gap in this paper by constructing
additional symmetries for the two  new sub-hierarchies and their
actions on eigenfunctions and adjoint eigenfunctions. The basic idea
is to revise the additional symmetry flows of CKP and BKP
hierarchies such that the new symmetry flows $\partial_k^*$ will
preserve the form of Lax operator of the cKP hierarchy. This will be
realized by introducing two vital operators $Y_k$ in sequel.

The paper is organized as follows. We first recall some basic
results of the KP hierarchy, CKP hierarchy, BKP hierarchy and the
constrained KP hierarchy in Section 2.  The main results are stated
and proved in Sections 3 and 4, which are the additional symmetries
and its action on the eigenfunction and adjoint eigenfunction of the
cCKP and  cBKP hierarchies, and we further show that acting on the
space of the wave operator, $\partial_k^*$ forms new centerless
$W^{cC}_{1+\infty}$ and $W^{cB}_{1+\infty}$-subalgebra of centerless
$W_{1+\infty}$ respectively. Section 5 is devoted to conclusions and
discussions.

\section{background on KP hierarchy}

In this section we first give a brief introduction of KP hierarchy
based on \cite{dkjm,dl1}. Let the pseudo-differential operator
\begin{equation}\label{KPlaxoperator}
L=\partial +u_1\partial^{-1}+u_2\partial^{-2}+
u_3\partial^{-3}+\cdots
\end{equation}
be a Lax operator of the KP hierarchy, which is described by the
associated Lax equations
\begin{equation}\label{KPhierarchy}
\dfrac{\partial L}{\partial {t_n}}=[B_n, L],\ n=1,2,3, \cdots,
\end{equation}
where $B_n=(L^n)_+=\sum\limits_{k=0}^n a_k\partial^k$ denotes the
non-negative powers of $\partial $ in $L^n$, $\partial
=\partial/\partial x$, $u_i=u_i(x=t_1,t_2,t_3,\cdots)$. The other
notation $L^n_{-}=L^n-L^n_+$ will be used in the paper. The Lax
operator $L$ in eq.(\ref{KPlaxoperator}) can be generated by
dressing operator $S=1+ \sum_{k=1}^{\infty}s_k \partial^{-k}$ in the
following way
\begin{equation}
L=S  \partial  S^{-1}.
\end{equation}
The dressing operator $S$ satisfies Sato equation
\begin{equation}
\dfrac{\partial S}{\partial t_n}=-(L^n)_-S, \quad n=1,2, 3, \cdots.
\end{equation}
The wave function $w(t,z)$ of the KP hierarchy  is defined by
\begin{equation}
w(t,z)=S e^{\xi(t,z)},
\end{equation}
 where
$\xi(t,z)=t_1z+t_2z^2+t_3z^3+\cdots+t_nz^n\cdots.$ The wave function
satisfies the equations
\begin{equation}
L^k w(t,z)=z^k w(t,z), \dfrac{\partial}{\partial t_n}w=B_nw, k \in
\mathbb{Z}, n \in \mathbb{N}.
\end{equation}
Moreover $w(t,z)$ has a very simple expression by $\tau$ function of
the KP hierarchy
\begin{equation}\label{KPtaufunction}
w(t,z)= \frac{\tau(t_1 - \frac{1}{z}, t_2 - \frac{1}{2z ^2}, t_3 -
\frac{1}{3z ^3}, \cdots)}{\tau(t_1, t_2, t_3, \cdots)}e^{\xi(t,z)}.
\end{equation}
Beside the existence of the Lax operator, wave function, $\tau$
function for the KP hierarchy, another important property is the
Zakharov-Shabat equation and
 associated linear  equation. In other words, KP
 hierarchy also has an alternative expression, i.e.,
\begin{gather}\label{zsforkp}
  \frac{\partial B_m}{\partial t_n}-\frac{\partial B_n}{\partial
  t_m}+[B_m,B_n]=0, \qquad m,n=1,2,3,\cdots.
  \end{gather}
The eigenfunction $\Phi=\Phi(t_1,t_2,t_3,\cdots)$ and adjoint
eigenfunction $\Psi=\Psi( t_1,t_2,t_3,\cdots  )$ of KP hierarchy
associated to (\ref{zsforkp}) are defined by
\begin{equation}
\dfrac{\partial \Phi}{\partial t_n}=(B_n\Phi), \  \dfrac{\partial
\Psi}{\partial t_n}=-(B_n^{*}\Psi),
\end{equation}
where the symbol $*$ is the formal adjoint operation. For an
arbitrary pseudo-differential operator $P=\sum\limits_i p_i
\partial^i$,  $P^*=\sum\limits_i (-1)^i\partial^i  p_i$, and
$(AB)^*=B^*A^*$ for  pseudo-differential operators $A$ and $B$. For
example, $\partial^*=-\partial$, $(\partial^{-1})^*=-\partial^{-1}$.

To construct additional symmetries associated with the KP hierarchy,
let us introduce  $\Gamma$ and  Orlov-Shulman's
 $M$ operator as $ \Gamma = \sum\limits_{i=1}^\infty i t_i\partial^{i-1} $ and $ M=S
\Gamma S^{-1}. $ Dressing $[\partial _k - \partial ^k, \Gamma] = 0$
gives
 $
[\partial_k - B_k, M] = 0, $ i.e.
\begin{equation}\label{KPmoperator1}
\partial _kM = [B_k, M],
\end{equation}
 and then
\begin{equation}
\partial _k (M^mL^n)=[B_k,M^mL^n].
\end{equation}
Thus, the additional flows are defined by
\begin{equation}
\dfrac{\partial S}{\partial t_{m,n}^*}=-(M^mL^n)_-S,
\end{equation}
or equivalently
\begin{equation}
\dfrac{\partial L}{\partial t_{m,n}^*}=-[(M^mL^n)_-,L].
\end{equation}
Here $(M^mL^n)_-$ serves as the generator of the additional flows
along the new time variables $t_{m,n}^*$.  The additional flows
${\partial_{m,n}^*}=\dfrac{\partial }{\partial t_{m,n}^*}$ commute
with the hierarchy, i.e. $[\partial_{m,n}^*,\partial_k]=0$ but do
not commute with each other, they indeed determine symmetries.
Therefore these flows $\partial_{m,n}^*$  are also called additional
symmetry flows, and they forms a centerless $W_{1+\infty}$
algebra\cite{asv2,dl4} acting on the spaces of the Lax operator $L$
and wave operator $S$. \vspace{8pt}

The CKP hierarchy is a reduction of the KP hierarchy through the
constraint on  $L$ given by eq.(\ref{KPlaxoperator})  as
\begin{equation}
L^*=-L,
\end{equation}
then $L$ is called the Lax operator of the CKP hierarchy, and the
associated  Lax equation of the CKP hierarchy is
\begin{equation}
\dfrac{\partial L}{\partial t_n}=[B_n, L], n=1, 3, 5, \cdots.
\end{equation}
which compresses all even flows,  i.e. the Lax equation of the CKP
hierarchy has only odd flows. Additional symmetry flows
\cite{he2007} for the  CKP hierarchy are defined as
\begin{equation}
\dfrac{\partial L}{\partial{t^*_{m,l}}}=-[(A_{m,l})_{-}, L],
\end{equation}
where $A_{m,l}$ for the  CKP hierarchy should satisfy
\begin{equation}\label{CKPconstraintonAml}
A_{m,l}^*=-A_{m,l},
\end{equation}
we can assume $A_{m,l}$  be
\begin{equation}\label{CKPadditionalsymmeterygenerator}
A_{m,l}=M^mL^l-(-1)^lL^lM^m,
\end{equation}
where
\begin{equation}\label{CKPMoperator}
M=S(\sum\limits_{i=0}^\infty (2i+1) t_{2i+1}\partial^{2i}) S^{-1}.
\end{equation}
Acting on the space of the wave operator $S$, $\partial_{m,l}^*$ of
the CKP hierarchy forms a new centerless $W^C_{1+\infty}$-
subalgebra of centerless $W_{1+\infty}$.

If Lax operator $L$ given by eq.(\ref{KPlaxoperator})  satisfies
\begin{equation}
 L^*=- \partial L \partial^{-1},
\end{equation}
then $L$ is called the Lax operator of the BKP hierarchy,  the Lax
equation of the BKP hierarchy also has only odd flows
\begin{equation}
\dfrac{\partial L}{\partial t_n}=[B_n, L], n=1, 3, 5, \cdots.
\end{equation}

Additional symmetry flows \cite{Takasaki1993,tu1} for the  BKP
hierarchy are given by
\begin{equation}
\dfrac{\partial L}{\partial{t^*_{m,l}}}=-[(A_{m,l})_{-}, L],
\end{equation}
where $A_{m,l}$ for the  BKP hierarchy should satisfy the constraint
equation
\begin{equation}\label{BKPconstraintonAml}
A_{m,l}^*=-\partial A_{m,l} \partial^{-1},
\end{equation}
then $A_{m,l}$ for the BKP hierarchy could  be chosen as
\begin{equation}\label{CKPadditionalsymmeterygenerator}
A_{m,l}=M^mL^l-(-1)^lL^{l-1}M^mL,
\end{equation}
where
\begin{equation}\label{BKPMoperator}
M=S(\sum\limits_{i=0}^\infty (2i+1) t_{2i+1}\partial^{2i}) S^{-1}.
\end{equation}
Acting on the space of the wave operator $S$, $\partial_{m,l}^*$ of
the BKP hierarchy forms another new centerless $W^B_{1+\infty}$ -
subalgebra of centerless $W_{1+\infty}$.

We now turn to the cKP hierarchy
\cite{kss,chengyiliyishenpla1991,chengyiliyishenjmp1995}.   The Lax
operator $L$ of  the constrained KP hierarchy is given by
\begin{equation}
L=\partial +\sum _{i=1}^m \Phi_i \partial ^{-i}\Psi _i,
\end{equation}
where $\Phi_i$ ($ \Psi_i$) is the (adjoint) eigenfunctions of this
hierarchy,  and the corresponding Lax equation  is formulated as
\begin{equation}
\dfrac{\partial L}{\partial t_n}=[B_n, L], n=1, 2, 3, \cdots.
\end{equation}
In particular, we stress that eigenfunctions and adjoint
eigenfunctions \{$\Phi_i, \Psi_i$\} are important dynamical
variables in cKP hierarchy,  so it is necessary to find the action
of the correct  additional symmetry flows on them. Here the correct
flow $\partial_\tau$  means its action on $L$ of the cKP hierarchy
should be the form of\cite{aratyn1997}
\begin{equation}\label{partialtaucKP}
(\partial_\tau L)_-=\sum_{i=1}^{m}( \tilde{A} \partial
^{-1}\Psi_i+\Phi_i\partial^{-1}\tilde{B}),
\end{equation}
which will result in its action on ($\partial_\tau
\Phi_i=\tilde{A}$) and ($\partial_\tau \Psi_i=\tilde{B}$). However,
in general, the additional symmetry flows of the KP hierarchy acting
on $L$ of the cKP hierarchy are not  the form of
eq.(\ref{partialtaucKP}). In other words, these flows do not
preserve the form of the Lax operator of the cKP hierarchy, the fact
shows additional symmetry of the KP hierarchy is not consistent with
cKP reduction condition automatically.  Therefore, the additional
symmetry flows have to be revised according to the analysis above,
and then the correct additional symmetry flows of the cKP hierarchy
are given by\cite{aratyn1997}
\begin{equation}
\partial^*_{k}L=[-(ML^k)_{-}+X_{k-1}, L],
k=0,1,2,3,\cdots,
\end{equation}
where $$X_{k-1}=0,  k=0,1,2$$ and
\begin{equation}
X_{k-1}=\sum _{i=1}^m \sum
_{j=0}^{k-2}(j-\dfrac{1}{2}(k-2))L^{k-2-j}(\Phi_i)
\partial ^{-1}(L^*)^j(\Psi_i),  k \geq 3.
\end{equation}
Furthermore\cite{aratyn1997},
\begin{align}
\partial_k^*\Phi_i &=\frac{k}{2}L^{k-1}(\Phi_i)+X_{k-1}(\Phi_i)+(A_{1,k})_+(\Phi_i),\\
\partial_k^*\Psi_i &=\frac{k}{2}L^{*(k-1)}(\Psi_i)-X_{k-1}^*(\Psi_i)-(A_{1,k})_+^*(\Psi_i).
\end{align}

\section{Additional symmetries of constrained  CKP hierarchy}

Let us consider the constrained CKP hierarchy, which is the C-type
sub-hierarchy of cKP hierarchy. The Lax operator $L$ of the cCKP
hierarchy \cite{Loris1999} is given by
\begin{equation}\label{CKPLaxoperator}
L=\partial+\sum_{i=1}^{m}(q_i\partial ^{-1}r_i+r_i\partial^{-1}q_i),
\end{equation}
where  $q_i$ and $r_i$ are  eigenfunctions. The corresponding Lax
equation of the cCKP hierarchy is defined by

\begin{equation}
\dfrac{\partial L}{\partial t_n}=[B_n, L], n=1, 3, 5, \cdots.
\end{equation}

For $k=1,3,5,\cdots$, we  first try to calculate the  original
additional symmetry flows of the CKP hierarchy as

\begin{equation}\label{cckpadd}
\dfrac{\partial L}{\partial{t^*_{1,k}}}=-[(A_{1,k})_{-}, L],
\end{equation}
where $ A_{1,k}=ML^k-(-1)^kL^kM.$ Thus

\begin{equation} \label{addsymmetryoperatorcCKP}
(\dfrac{\partial
L}{\partial{t^*_{1,k}}})_-=[(A_{1,k})_+,L]_-+2(L^k)_-.
\end{equation}
In order to get its action on $q_i$ and $r_i$, we need the form of
$(L^k)_-$.

\vspace{6pt}

{\bf Lemma 3.1.} The Lax operator  $L$ of the constrained CKP
hierarchy  given by eq.(\ref{CKPLaxoperator}) satisfies the relation
\begin{equation}\label{cCKPLKNEG}
(L^k)_-=\sum _{i=1}^m \sum _{j=0}^{k-1}(L^{k-j-1}(q_i)
\partial ^{-1}(L^*)^j(r_i)+L^{k-j-1}(r_i)
\partial ^{-1}(L^*)^j(q_i)), k=1,3,5,\cdots
\end{equation}
where
$$L(q_i)= L_+(q_i)+\sum_{j=1}^{m}(q_j\partial_x ^{-1}(r_jq_i)+r_j \partial _x^{-1}(q_jq_i)).$$
However,  $(L^k)_-$ is not in the form of
\begin{equation}\label{CKPconform}
(\partial_\tau L)_-=\sum_{i=1}^{m}((\partial_\tau q_i)\partial
^{-1}r_i+(\partial_\tau r_i)\partial^{-1}q_i+q_i\partial
^{-1}(\partial_\tau r_i)+r_i\partial^{-1}(\partial_\tau q_i)),
\end{equation}
as we expected for a correct flows. Here correctness means we can
get the actions on $\partial_\tau q_i$ and $\partial_\tau r_i$ from
eq.(\ref{CKPconform}). This shows that
eq.(\ref{addsymmetryoperatorcCKP})
 can not imply its action on eigenfunctions $\partial^*_{1,k}q_i$ and $\partial^*_{1,k}r_i$.

Therefore, in order to get correct additional flows of the cCKP
hierarchy, we revise eq.(\ref{cckpadd}) and then define new flows by
\begin{equation} \label{CKPaddsymm}
\partial _k^* L=[-(A_{1,k})_{-}+Y_k,
L],k=1,3,5,\cdots,
\end{equation}
where $A_{1,k}=ML^k-(-1)^kL^kM$ and $Y_k$ is introduced  such that
the left hand side of eq.(\ref{CKPaddsymm}) will be the form of
eq.(\ref{CKPconform}).  Next, we shall prove new flows
$\partial_k^*$ are correct additional symmetry flows of the cCKP
hierarchy. First of all, $Y_k$ is discussed, which is crucial to our
purpose.

\vspace{6pt}

{\bf Lemma 3.2.} For the cCKP hierarchy, the CKP reduction
condition infers a constraint on $Y_k$,
\begin{equation}\label{cCKPYKcondition}
Y_k^*=-Y_k, \ k=1,3,5,\cdots.
\end{equation}
Proof. The action of the additional flows $\partial _k^*$ on
the adjoint Lax operator $L^*$ of the constrained CKP hierarchy can
be obtained  by two different ways. We first computet it by a formal
adjoint operation on both sides of eq. (\ref{CKPaddsymm}), i.e.
\begin{equation} \label{CKPaddsymmadj1}
\partial _k^* L^*=([-(A_{1,k})_{-}+Y_k,
L])^*=[L^*,-(A_{1,k})_{-}^*+Y_k^*],k=1,3,5,\cdots.
\end{equation}
The another way is to do a derivative with respect to $t_k^*$ on
$L^*$ and use CKP reduction condition $L^{*}=-L$, then
\begin{equation} \label{CKPaddsymmadj2}
\partial _k^* L^*=-\partial _k^* L=-[-(A_{1,k})_{-}+Y_k,
L]=[L^*,(A_{1,k})_--Y_k],k=1,3,5,\cdots.
\end{equation}
Comparing eq.(\ref{CKPaddsymmadj1}) and eq.(\ref{CKPaddsymmadj2}),
we have
\begin{equation*}
Y_k^*=-Y_k, \ k=1,3,5,\cdots,
\end{equation*}
with the help of $A_{1,k}^*=-A_{1,k}$. \hfill $\square$

Thus we set $Y_k$ as
\begin{align}
Y_1 & =0,  \\  Y_k & =\sum _{i=1}^m \sum
_{j=0}^{k-2}(2j-(k-2))(L^{k-2-j}(q_i)
\partial ^{-1}(L^*)^j(r_i)+L^{k-2-j}(r_i)
\partial ^{-1}(L^*)^j(q_i)),  k \geq 3, \label{CKPYK}
\end{align}
and shall show  $Y_k$ satisfy the constraint given by
eq.(\ref{cCKPYKcondition}) .

\vspace{6pt}

{\bf Lemma 3.3.} 
\begin{equation}
Y_k=-Y_k^*.
\end{equation}

Proof. $ Y_1^*=Y_1=0 $ is obvious, for $k=3,5,7,\cdots$, we
have
\begin{align*}
Y_k^* & =\sum _{i=1}^m \sum _{j=0}^{k-2}(2j-(k-2))(-(L^*)^j(r_i)
\partial ^{-1} L^{k-2-j}(q_i) - (L^*)^j(q_i)
\partial ^{-1}L^{k-2-j}(r_i)) \\
& =\sum _{i=1}^m \sum _{j=0}^{k-2}(2j-(k-2))(-(-1)^jL^j(r_i)
\partial ^{-1}(-1)^{k-2-j} (L^*)^{k-2-j}(q_i) \\
& - (-1)^jL^j(q_i)
\partial ^{-1}(-1)^{k-2-j} (L^*)^{k-2-j}(r_i)) \\
& =\sum _{i=1}^m \sum _{j=0}^{k-2}(2j-(k-2))(L^j(r_i)
\partial ^{-1}(L^*)^{k-2-j}(q_i) +L^j(q_i)
\partial ^{-1} (L^*)^{k-2-j}(r_i)) \\
&=-\sum _{i=1}^m \sum _{l=0}^{k-2}(2l-(k-2))(L^{k-2-l}(q_i)
\partial ^{-1}(L^*)^l(r_i)+L^{k-2-l}(r_i)
\partial ^{-1}(L^*)^l(q_i)) \\
&=-Y_k
\end{align*}
In the fourth step we  let $l=k-2-j$. \hfill $\square$

 Furthermore, in order to calculate $[Y_k,L]$ in the
eq.(\ref{CKPaddsymm}), the following lemma is necessary.

\vspace{6pt}

{\bf Lemma 3.4.} For the Lax operator $L$ of  the cCKP
hierarchy in eq.(\ref{CKPLaxoperator})  and a pseudo-different
operator $X=\sum _{k=1}^lM_k\partial ^{-1}N_k$, the equation
\begin{align*}
[X,L]_- & =\sum _{k=1}^l(-L(M_k)
\partial ^{-1}N_k+M_k
\partial ^{-1}L^*(N_k)) \\
& +\sum_{i=1}^{m}(X(q_i)\partial ^{-1}r_i+X(r_i)\partial
^{-1}q_i-q_i\partial ^{-1}X^*(r_i)-r_i\partial ^{-1}X^*(q_i))
\end{align*}
holds.

\vspace{6pt}

Based on the above lemma, we have
$$\hspace{0cm}
[Y_k,L]_-=-2(L^k)_-+k\sum_{i=1}^{m}(L^{k-1}(q_i)
\partial ^{-1}r_i+L^{k-1}(r_i)
\partial ^{-1}q_i)$$
$$ \hspace{0.6cm}
+k\sum_{i=1}^{m}(q_i
\partial ^{-1}L^{*(k-1)}(r_i)+r_i
\partial ^{-1}L^{*(k-1)}(q_i))
$$
$$\hspace{-1cm}+\sum_{i=1}^{m}(Y_k(q_i)\partial ^{-1}r_i+Y_k(r_i)\partial ^{-1}q_i)$$
\begin{equation}\label{YkLcommutator}
\hspace{-1cm}-\sum_{i=1}^{m}(q_i\partial ^{-1}Y_k^*(r_i)+r_i\partial
^{-1}Y_k^*(q_i)).
\end{equation}
We are now in a position to calculate the explicit form of the right
hand side of the new additional flow give by eq.(\ref{CKPaddsymm}).

\vspace{6pt}

{\bf Theorem 3.1.}  The additional flows acting on the
 eigenfunction $q_i$ and  $r_i$ of the constrained  CKP hierarchy are
\begin{equation}
\partial _k^* q_i=kL^{k-1}(q_i)+Y_k(q_i)+(A_{1,k})_{+}(q_i) ,k=1,3,5,\cdots
\end{equation}
and
\begin{equation}
\partial _k^* r_i=kL^{k-1}(r_i)+Y_k(r_i)+(A_{1,k})_{+}(r_i) ,k=1,3,5,\cdots.
\end{equation}

Proof. From the revised definition of the additional symmetry
flows  in eq.(\ref{CKPaddsymm}) of the cCKP hierarchy, by a short
calculation, we have
$$
\partial _k^* L_-=[-(A_{1,k})_{-}+Y_k,L]_-=[(A_{1,k})_+,L]_-+2(L^k)_-+ [Y_k,L]_- .$$
Using eq.(\ref{YkLcommutator}) and  the technical identity
$[K,f\partial^{-1} g]_-=K(f)\partial^{-1}g-f\partial^{-1}K^*(g)$
(where $f,$ $g$ are arbitrary functions and $K  $ is a purely
differential operator), \ note that $k=1,3,5,\cdots,$ it is followed
by
\begin{align*}
\partial _k^* L_- &=\sum_{i=1}^{m}((kL^{k-1}(q_i)+Y_k(q_i)+(A_{1,k})_{+}(q_i))\partial
^{-1}r_i \\
&+(kL^{k-1}(r_i)+Y_k(r_i)+(A_{1,k})_{+}(r_i))\partial^{-1}q_i \\
&+q_i\partial ^{-1}(k(L^*)^{k-1}(r_i)-Y_k^*(r_i)-(A_{1,k})_{+}^*(r_i)) \\
&+r_i\partial^{-1}(k(L^*)^{k-1}(q_i)-Y_k^*(q_i)-(A_{1,k})_{+}^*(q_i))
\end{align*}
thus
$$\partial _k^* q_i=kL^{k-1}(q_i)+Y_k(q_i)+(A_{1,k})_{+}(q_i)=k(L^*)^{k-1}(q_i)-Y_k^*(q_i)-(A_{1,k})_{+}^*(q_i),$$
$\partial _k^* r_i$ is also obtained at the same time. \hfill
$\square$

\vspace{6pt}

{\bf Corollary 3.1.} The additional flows act on wave operator $S$ of the constrained CKP
hierarchy  as
\begin{equation}\label{CKPaddsymmonwave}
\partial _k^* S=(-(A_{1,k})_{-}+Y_k )S ,k=1,3,5,\cdots.
\end{equation}

\vspace{6pt}

{\bf Theorem 3.2.} The additional flows ${\partial_{k}^*}$
commute with the constrained CKP hierarchy flows
$\partial_{t_{2n+1}}=\dfrac{\partial }{\partial{t_{2n+1}}}$, i.e.
\begin{equation}
[{\partial_{k}^*}, \partial_{t_{2n+1}}]=0.
\end{equation}
Thus $\partial_k^*(k=1,3,5,\cdots)$ are indeed the additional
symmetry flows of the cCKP hierarchy.

Proof. The proof starts with  the definition
\begin{equation*}
[\partial_{k}^*, \partial_{t_{2n+1}}] S=\partial_{k}^*
(\partial_{t_{2n+1}}S)-
\partial_{t_{2n+1}} (\partial_{k}^*S),
\end{equation*}
and using the action of the additional flows, we get
\begin{eqnarray*}
[\partial_{k}^*,\partial_{t_{2n+1}}]S&=&
-\partial_{k}^*\left(L^{2n+1}_{-}S\right) +
\partial_{t_{2n+1}} \left(((A_{1,k})_{-}-Y_k)S \right)\\
&=&-(\partial_{k}^*L^{2n+1} )_{-}S- (L^{2n+1})_{-}(\partial_{k}^*S)+
\partial_{t_{2n+1}} ((A_{1,k})_{-}-Y_k)S \\
&+&((A_{1,k})_{-}-Y_k)(\partial_{t_{2n+1}}S).
\end{eqnarray*}

Taking eq.(\ref{CKPaddsymmonwave}) of Corollary 1 into the above
formula, it is not difficult to compute
\begin{eqnarray*}
[\partial_{k}^*,\partial_{t_{2n+1}}]S&=&[(A_{1,k})_{-}-Y_k,
L^{2n+1}]_{-}S+
(L^{2n+1})_{-}((A_{1,k})_{-}-Y_k)S\\
&+&[(L^{2n+1})_{+},(A_{1,k})_{-}-Y_k]_{-}S-((A_{1,k})_{-}-Y_k)(L^{2n+1})_{-}S\\
&=&[((A_{1,k})_{-}-Y_k), L^{2n+1}]_{-}S- [(A_{1,k})_{-}-Y_k,
L^{2n+1}_{+}]_{-}S+
[L^{2n+1}_{-},(A_{1,k})_{-}-Y_k]S\\
&=&[(A_{1,k})_{-}-Y_k, L^{2n+1}_{-}]_{-}S+
[L^{2n+1}_{-},(A_{1,k})_{-}-Y_k]S=0
\end{eqnarray*}

We have used the fact that $[L^{2n+1}_{+}, (A_{m,l})-Y_{k}]_{-}=
[L^{2n+1}_{+}, ((A_{1,k})_{-}-Y_k)]_{-}$ in the second step of the
above derivation, since $(Y_{k})_-=Y_{k}$ and $[L^{2n+1}_{+},
(A_{m,l})_{+}]_{-}=0$. The last equality holds by virtue of
$(P_{-})_{-}=P_{-}$ for arbitrary pseduo-differential operator $P$.
\hfill $\square$

\vspace{6pt}

 Taking into account of $Y_k^*=-Y_k, \ k=1,3,5,\cdots$,
we can give the next theorem by a straightforward and tedious
calculation.

\vspace{6pt}

{\bf Theorem 3.3.} Acting on the space of the wave operator $S$
of the constrained CKP hierarchy, $\partial_k^*$ forms a new
centerless $W^{cC}_{1+\infty}$-subalgebra of centerless
$W_{1+\infty}$.

\section{Additional
symmetries of constrained  BKP hierarchy}

We now turn to the case of the constrained  BKP hierarchy, this
follows in a similar way of the case of  the constrained  CKP
hierarchy. The Lax operator $L$ of the constrained BKP hierarchy
\cite{Loris1999JMP} is given by
\begin{equation}\label{BKPLaxoperator}
L=\partial+\sum_{i=1}^{m}(q_i\partial
^{-1}r_{i,x}-r_i\partial^{-1}q_{i,x}),
\end{equation}
where the $q_i$ and $r_i$  are eigenfunctions. The corresponding Lax
equation of the constrained BKP hierarchy is defined by
\begin{equation}
\dfrac{\partial L}{\partial t_n}=[B_n, L], n=1, 3, 5, \cdots.
\end{equation}
For $k=1,3,5,\cdots$, at first we calculate the  original additional
symmetry flows of the BKP hierarchy as
\begin{equation}\label{originaaddsymmt1k}
\dfrac{\partial L}{\partial{t^*_{1,k}}}=-[(A_{1,k})_{-}, L],
\end{equation}
where $ A_{1,k}=M^mL^l-(-1)^lL^{l-1}M^mL.$ Thus
\begin{equation}
(\dfrac{\partial
L}{\partial{t^*_{1,k}}})_-=[(A_{1,k})_+,L]_-+2(L^k)_-.
\end{equation}
For the cBKP hierarchy,  the action of one flow $\partial_\tau $ on
the eigenfunctions $(\partial_\tau q_{i})$ and $(\partial_\tau
r_{i})$ may be derived from
 its action on the $L$ if
\begin{equation}\label{BKPconform}
(\partial_\tau L)_-=\sum_{i=1}^{m}((\partial_\tau q_i)\partial
^{-1}r_{i,x}-(\partial_\tau r_i)\partial^{-1}q_{i,x}+q_i\partial
^{-1}(\partial_\tau r_{i,x})-r_i\partial^{-1}(\partial_\tau
q_{i,x})).
\end{equation}
At the same time, $(\partial_\tau q_{i,x})$ should be consistent
with $ (\partial_\tau q_{i})$, $(\partial_\tau r_{i,x})$ should be
consistent with $ (\partial_\tau r_{i})$. However, the following
lemma shows that $\partial_{t^*_{1,k}}$ is not the case due to the
form of $(L^k)_-$.

\vspace{6pt}

{\bf Lemma 4.1.} The Lax operator  $L$ of the constrained BKP
hierarchy  given by eq.(\ref{BKPLaxoperator}) satisfies the relation
\begin{equation}
(L^k)_-=\sum _{i=1}^m \sum _{j=0}^{k-1}(L^{k-j-1}(q_i)
\partial ^{-1}(L^*)^j(r_{i,x})-L^{k-j-1}(r_i)
\partial ^{-1}(L^*)^j(q_{i,x})), k=1,3,5,\cdots
\end{equation}
where
$$L(q_i)= L_+(q_i)+\sum_{j=1}^{m}(q_j\partial_x ^{-1}(r_{j,x}q_i)-r_j \partial_x^{-1}(q_{j,x}q_i)).$$

According to the  analysis above, we may revise the flows in
eq.(\ref{originaaddsymmt1k}) to define a correct additional symmetry
flows of the cBKP hierarchy. Similar to the case of cCKP hierarchy,
we define a new additional flow
\begin{equation} \label{BKPaddsymm}
\partial _k^* L=[-(A_{1,k})_{-}+Y_k,
L],k=1,3,5,\cdots,
\end{equation}
where $A_{1,k}=M^mL^l-(-1)^lL^{l-1}M^mL$ is the generator of the
additional symmetry of the BKP hierarchy. We shall show in this
section that $\partial_k^*$ is a correct additional symmetry flow
for the cBKP hierarchy. First of all, by a similar discussion as the
case of the cCKP hierarchy, we have the  following lemma.

\vspace{6pt} 

{\bf Lemma 4.2.} For the cBKP hierarchy, the BKP
reduction condition infers a constraint on $Y_k$ as
\begin{equation}
Y_k^*=-\partial Y_k \partial^{-1}, \ k=1,3,5,\cdots
\end{equation}

Thus we can set
\begin{align}
Y_1 & =0,\\
Y_k & =\sum _{i=1}^m \sum _{j=0}^{k-2}(2j-(k-2))(L^{k-2-j}(q_i)
\partial ^{-1}(L^*)^j(r_{i,x})-L^{k-2-j}(r_i)
\partial ^{-1}(L^*)^j(q_{i,x})), k \geq 3.
\end{align}

The following lemma shows that  $\{Y_k, \ k=1,3,5,\cdots\}$  satisfy
the property we need.

\vspace{6pt}

{\bf Lemma 4.3.}
\begin{equation}
Y_k^*=-\partial Y_k \partial^{-1}.
\end{equation}

Proof.  $ Y_1^*=Y_1=0 $ is trivial. For $k=3,5,7,\cdots$, we have
\begin{align*}
\partial Y_k \partial^{-1} & =\sum _{i=1}^m \sum _{j=0}^{k-2}(2j-(k-2))(\partial L^{k-2-j}(q_i)
\partial ^{-1}(L^*)^j(r_{i,x})\partial^{-1}-\partial L^{k-2-j}(r_i)
\partial ^{-1}(L^*)^j(q_{i,x})\partial^{-1}) \\
& =\sum _{i=1}^m \sum
_{j=0}^{k-2}(2j-(k-2))((-1)^{k-2-j}(L^*)^{k-2-j}(q_{i,x})(-1)^j(L^j(r_i)\partial
^{-1}-\partial ^{-1}L^j(r_i)) \\
& -(-1)^{k-2-j}(-L^*)^{k-2-j}(r_{i,x})(-1)^j(L^j(q_i)\partial
^{-1}-\partial ^{-1}L^j(q_i))) \\
& =-\sum _{i=1}^m \sum
_{j=0}^{k-2}(2j-(k-2))((L^*)^{k-2-j}(q_{i,x})L^j(r_i)\partial
^{-1}-(L^*)^{k-2-j}(q_{i,x})\partial ^{-1}L^j(r_i) \\
& -(L^*)^{k-2-j}(r_{i,x})L^j(q_i)\partial
^{-1}+(L^*)^{k-2-j}(r_{i,x})\partial ^{-1}L^j(q_i)) \\
& =-\sum _{i=1}^m \sum _{l=0}^{k-2}(2l-(k-2))(-(L^*)^l(r_{i,x})
\partial ^{-1} L^{k-2-l}(q_i) +
(L^*)^l(q_{i,x})\partial ^{-1}L^{k-2-l}(r_i))\\
& -\sum _{i=1}^m \sum _{j=0}^{k-2}(2j-(k-2))(-1)^{k-2-j}(
\partial L^{k-2-j}(q_i)L^j(r_i)\partial ^{-1}-
\partial L^{k-2-j}(r_i)L^j(q_i)\partial ^{-1})
\\
& =-Y_k^*
\end{align*}
we have used the identity $
\partial^{-1}f_x\partial^{-1}=f\partial^{-1}-\partial^{-1}f$ as
$f=(L^*)^j(r_{i,x})$ and $f=(L^*)^j(q_{i,x})$ in  the derivation.
 \hfill $\square$

In order to get the explicit form of the right hand side of
eq.(\ref{BKPaddsymm}), the following lemma is necessary.

\vspace{6pt}

{\bf Lemma 4.4.} For the Lax operator $L$ of the cBKP hierarchy
and $X=\sum _{k=1}^lM_k\partial ^{-1}N_k$,
\begin{align*}
[X,L]_- &=\sum _{k=1}^l(-L(M_k)
\partial ^{-1}N_k+M_k
\partial ^{-1}L^*(N_k))\\
&
 +\sum_{i=1}^{m}(X(q_i)\partial ^{-1}r_{i,x}-X(r_i)\partial
^{-1}q_{i,x}-q_i\partial ^{-1}X^*(r_{i,x})+r_i\partial
^{-1}X^*(q_{i,x}))
\end{align*} holds.

\vspace{6pt}

 Applying  the above lemma we conclude that
$$\hspace{0cm}
[Y_k,L]_-=-2(L^k)_-+k\sum_{i=1}^{m}(L^{k-1}(q_i)
\partial ^{-1}r_{i,x}-L^{k-1}(r_i)
\partial ^{-1}q_{i,x})$$
$$ \hspace{0.8cm}
+k\sum_{i=1}^{m}(q_i
\partial ^{-1}L^{*(k-1)}(r_{i,x})-r_i
\partial ^{-1}L^{*(k-1)}(q_{i,x}))
$$
$$\hspace{-1.0cm}+\sum_{i=1}^{m}(Y_k(q_i)\partial ^{-1}r_{i,x}-Y_k(r_i)\partial ^{-1}q_{i,x})$$
$$\hspace{-0.7cm}-\sum_{i=1}^{m}(-q_i\partial ^{-1}Y_k^*(r_{i,x})+r_i\partial ^{-1}Y_k^*(q_{i,x})).$$

\vspace{6pt}

{\bf Theorem 4.1.} For the cBKP hierarchy, the additional flows defined by eq.(\ref{BKPaddsymm}) acting on the
 eigenfunction $q_i$ and $r_i$  are
\begin{align}
& \partial _k^* q_i \ \ =kL^{k-1}(q_i)+Y_k(q_i)-(A_{1,k})_{+}(q_i), \\
& \partial _k^* r_i \ \ =kL^{k-1}(r_i)+Y_k(r_i)+(A_{1,k})_{+}(r_i), \\
& \partial _k^*
q_{i,x}=-kL^{*(k-1)}(q_{i,x})-Y_k^*(q_{i,x})+(A_{1,k})_{+}^*(q_{i,x}),
\\
& \partial _k^* r_{i,x}
=-kL^{*(k-1)}(r_{i,x})-Y_k^*(r_{i,x})-(A_{1,k})_{+}^*(r_{i,x})
,k=1,3,5,\cdots.
\end{align}
Proof. From the revised definition of the additional flows  in
eq.(\ref{BKPaddsymm}) of the cBKP hierarchy, by a short calculation,
we have
\begin{align*}
\partial _k^* L_- &=[-(A_{1,k})_{-}+Y_k,L]_- \\
& =[(A_{1,k})_+,L]_-+2(L^k)_-+ [Y_k,L]_- \\
& =\sum_{i=1}^{m}(kL^{k-1}(q_i)+Y_k(q_i)-(A_{1,k})_{+}(q_i))\partial
^{-1}r_{i,x} \\
&-(kL^{k-1}(r_i)+Y_k(r_i)+(A_{1,k})_{+}(r_i))\partial^{-1}q_{i,x} \\
&+q_i\partial ^{-1}(-kL^{*(k-1)}(r_{i,x})-Y_k^*(r_{i,x})-(A_{1,k})_{+}^*(r_{i,x})) \\
&-r_i\partial^{-1}(-kL^{*(k-1)}(q_{i,x})-Y_k^*(q_{i,x})+(A_{1,k})_{+}^*(q_{i,x}))
\end{align*}
thus
$$\partial _k^* q_i \ \ =kL^{k-1}(q_i)+Y_k(q_i)-(A_{1,k})_{+}(q_i),$$
 the other three identities  could be also obtained in the same way. \hfill
$\square$

\vspace{6pt}

{\bf Remark 4.1.}  By a simple calculation,  $\partial _k^*
q_{i,x}$ and $ \partial _k^* r_{i,x}$ are not essential and
necessary, because it can be obtained from  $\partial _k^* q_{i}$
and $\partial _k^* r_{i}$, respectively.

\vspace{6pt}

{\bf Corollary 4.1.} The additional flows act on wave operator $S$ of the constrained BKP
hierarchy  as
\begin{equation}
\partial _k^* S=(-(A_{1,k})_{-}+Y_k )S ,k=1,3,5,\cdots.
\end{equation}

\vspace{6pt}

 {\bf Theorem 4.2.} The additional flows ${\partial_{k}^*}$ commute with constrained BKP
hierarchy $\dfrac{\partial }{\partial{t_{2n+1}}}$, \\ i.e.
\begin{equation}
[{\partial_{k}^*}, \partial_{t_{2n+1}}]=0.
\end{equation}

This theorem shows  $\partial_k^*(k=1,3,5,\cdots)$ are indeed the
additional symmetry flows of the cBKP hierarchy.

Using the identity $Y_k^*=-\partial Y_k \partial^{-1}, \
k=1,3,5,\cdots$, we can present the next theorem omitting the proof.

\vspace{6pt}

{\bf Theorem 4.3.} Acting on the space of the wave operator $S$ of the constrained BKP
hierarchy, $\partial_k^*$ forms new centerless
$W^{cB}_{1+\infty}$-subalgebra of centerless $W_{1+\infty}$.

\section{Conclusions and Discussions}
To summarize, we have constructed the additional symmetries of the
constrained CKP hierarchy in eq.(\ref{CKPaddsymm}) and theorem 3.2.
The additional flows action on the eigenfunction  of the cCKP
hierarchy are given in theorem 3.1.  Acting on the space of the wave
operator $S$ of the cCKP hierarchy, $\partial_k^*$ forms a new
centerless $W^{cC}_{1+\infty}$-subalgebra of centerless
$W_{1+\infty}$ algebra in theorem 3.3. Similarly, the conclusions
for the constrained BKP hierarchy are obtained  using the analogous
technique in the case of constrained CKP hierarchy. The main results
of the cBKP hierarchy are presented in eq.(\ref{BKPaddsymm}),
theorem 4.1, 4.2 and 4.3. Our results show that the
 cCKP and cBKP hierarchies have, indeed, some different properties for additional symmetry  comparing with the KP, BKP, CKP and
constrained KP hierarchies. For
 example, the definitions of   additional symmetry flows for the  cCKP and cBKP
 hierarchies are different from the above  hierarchies, the revised operators
  ${Y_k}$  of the  cCKP and cBKP
 hierarchies are different from the corresponding operator $X_k$ of the cKP  hierarchy given in
 \cite{aratyn1997}. Moreover, we also would like to point out that  ${Y_k}$  of the  cCKP and
 cBKP hierarchies are distinct. We can use  the similar technique without essential difficulty to
 construct additional symmetries of the case as $L^k=B_k+\sum q_i\partial^{-1}r_i+r_i\partial^{-1} q_i,
 k>1$, the results will be given in the future.

{\bf Acknowledgments} 
This work was supported by the National Natural
Science Foundation of China (Grant No.10971109, 10971209), and the
Program for New Century Excellent Talents in University (Grant
No.NCET-08-0515).


\end{document}